# Multi-Task Learning for Few-Shot Online Adaptation under Signal Temporal Logic Specifications

Andres Arias, *Student Member, IEEE*, Chuangchuang Sun, *Member, IEEE*

*Abstract*— **Multi-task learning (MTL) seeks to improve the generalized performance of learning specific tasks, exploiting useful information incorporated in related tasks. As a promising area, this paper studies an MTL-based control approach considering Signal Temporal Logic (STL). Task compliance is measured via the Robustness Degree (RD) which is computed by using the STL semantics. A suitable methodology is provided to solve the learning and testing stages, with an appropriate treatment of the non-convex terms in the quadratic objective function and using Sequential Convex Programming based on trust region update. In the learning stage, an ensemble of tasks is generated from deterministic goals to obtain a strong initializer for the testing stage, where related tasks are solved with a larger impact of perturbation. The methodology demonstrates to be robust in two dynamical systems showing results that meet the task specifications in a few shots for the testing stage, even for highly perturbed tasks.**

## I. Introduction

IN the learning process, developing a new skill is based on prior knowledge obtained with other related skills. This principle leads to learn multiple tasks simultaneously, that is compared with the mechanism used by the humans to learn via transferable skills. Knowledge transfer mechanism allows an individual to learn new concepts by performing few examples, framed into the Few-Shot Learning. Motivated by humans ability to learn, Multi task learning (MTL) aims to learn multiple tasks simultaneously such that the experience obtained when developing one task, gathers useful information that can be used to generalize the performance for other related tasks [1]. MTL results convenient for situations in which data observation is limited, since a good learner is often trained by using a noticeable amount of data instances [2]. This issue of data sparsity is particularly in applications where more manual labor to label data is required, i.e., medical image analysis, speech recognition and natural language processing [3]. Additionally, MTL not only improves the performance of specific tasks, but also increases the adversarial robustness, decreasing the model vulnerabilities under adversarial attacks. Significant progress has placed during the last two decades in the MTL area, furthermore its roots are originated in psychology and cognitive science [4]. Early contributions are presented in [5], defining MTL as an inductive transfer mechanism to improve the generalization performance, by leveraging the domain-specific information contained in training signals of related tasks. Training signals process is developed in parallel using a shared representation. Different approaches have been proposed for MTL, from multi-objective optimization [6], via dynamic task prioritization [7], to task-dependent weighting [8].

Conversely, Signal Temporal Logic (STL) is popular for temporal tasks in autonomous systems due to its expressiveness and natural language similarity. The task is in terms of one or several specifications with temporal parameters. The symbolic control problem is composed of a continuous-time signal trajectory that meets the specifications set forth by the task, providing a quantitative notion of time and space [9]. Several studies are conducted on the STL application in the optimal control context. This encompasses robust control, control barrier function and control Lyapunov function, applied in energy management systems, temperature control, autonomous driving and trajectory control for unmanned ground and aerial vehicles. In [10] the signal Robustness Degree (RD) subject to the system dynamics is maximized, using smooth approximations of STL semantics. The methodology is applied in the heating, ventilation and air conditioning building control. Similarly, in [11] a new robustness under-approximation is developed, based on arithmetic and geometric means that makes the positive robustness a sufficient condition to meet the specification. STL has been extended to time-varying Control Barrier Function (CBF) and its temporal properties within STL tasks satisfaction. A control law is derived so that the state variables solution with an initial condition meets the specification. This approach is useful, specially in multi-robot systems, where some methods do not scale computationally [12]. The same authors in [12], propose in [13] a methodology to construct a CBF for a fragment of STL tasks by solving an optimization problem, applied in dynamically coupled multi-agent systems. Other approaches are within learning a neural network controller to satisfy a STL specification. The parameters of the CBF are obtained through training, substantially decreasing the conservativeness [14]. Considering the trajectory planning for continuous linear systems, in [15] a mixed integer quadratic program is formulated with constraints obtained from the STL specifications with linear predicates

Andres Arias and Chuangchuang Sun are with the Department of Aerospace Engineering, Mississippi State University, Starkville, MS 39762 USA (e-mail: aa2935@msstate.edu; csun@ae.msstate.edu).

and CBF, to synthetize a discrete sequence of controls. In [16] the STL specifications are over nonlinear state constraints using High Order Control Lyapunov Barrier Functions. This leads to develop controllers for system stabilization inside a set within a specified time, while guaranteeing the system to remain in the set.

In this article, a MTL control for problems with STL specifications is addressed, using STL semantics and RD maximization. Due to the non-convex terms that arise from the STL semantics, such as, the log-sum exponential approximation for the min and max operators, Sequential Convex Programming (SCP) algorithm is used to solve the non-convex problem, considering the system dynamics and the trust region as constraints. Two stages are in the methodology: learning stage in which the SCP finds a feasible solution of the optimal problem under tasks perturbed in STL specifications, given that the RD of the final solution must be positive for task compliance. Then in test stage, other related tasks are generated by applying a larger perturbation compared with learning stage. The solution obtained in learning stage is the initializer for testing stage, solving the new perturbed tasks in few shots of the SCP. To the best of our knowledge, there have been no attempts for MTL under STL, since optimal control problems STL based can be tackled with MTL principles.

The rest of this article is organized as follows. Section II presents the mathematical formulation, considering the optimal control problem and how the STL framework introduces in RD. In Section III the problem formulation and solution methodology is explained, within the SCP algorithm and MTL learning and testing stages. The results of two case studies are addressed in Section IV. Finally, some concluding remarks are presented in Section V.

## II. PRELIMINARIES AND MATHEMATICAL FORMULATION

Let $x \in R^n$ and $u \in U \subseteq R^m$ be the state and control input variables of a discrete-time system presented in (1):

$$x_{k+1} = f(x_k, u_k) \quad (1)$$

where $f(x_k, u_k)$ represents the discretized system dynamics. Initial condition $x(t_0)$ is considered to solve (1) over a fixed and continuous-time interval $[t_0, t_{N_T}] \subset R_{\geq 0}$. Given the initial sate $x(t_0)$ and a control sequence $(u_0, u_1, ..., u_{N_T-1})$, a trajectory $(x_0, x_1, ..., x_{N_T})$ that satisfies (1) can be generated, and under this work is called signal.

### A. Signal Temporal Logic (STL) and Robustness Degree (RD)

STL is a formal language to describe a broad range of real-valued, temporal properties in cyber-physical systems [17]. An STL formula comprises predicates, logical connectives and temporal operators, using the atomic propositions presented in the Backus-Naur form, as shown in (2):

$$\phi := True \,|\, \mu \geq 0 \,|\sim \phi \,|\, \phi_1 \wedge \phi_2 \,|\, \phi_1 U_{[a,b]} \phi_2 \,|\, \phi_1 \Rightarrow \phi_2 | \quad (2)$$

where $\sim$ means "negation", $\wedge$ is the "And" operator, U means "Until", $[a,b]$ is the time interval from $a$ to $b$, and $\phi_1$ and $\phi_2$ are STL formulas. From the elementary propositions above, additional operators can be written, such as $\phi_1 \wedge \phi_2$, $F_{[a,b]}\phi$, $G_{[a,b]}\phi$ and $\phi_1 \Rightarrow \phi_2$, that stand for "Or", "Eventually", "Always" and "Infer" operators, respectively. The predicate $\mu$ is a function $h$ in terms of the signal $x$, which is expressed as $h(x) - c$. If $\mu \geq 0$, then $\mu$ is True, otherwise $\mu$ is False. Considering that $(x, t) \models \phi$ indicates if the signal $x$ satisfies $\phi$ at time $t$, the formula satisfaction is recursively computed in the formula semantics, as in (3).

$$\begin{aligned}
(x,t) &\models \mu &\Leftrightarrow\quad & h(x(t)) \geq 0 \\
(x,t) &\models\, \sim \phi &\Leftrightarrow\quad & \sim ((x,t) \Leftrightarrow \phi) \\
(x,t) &\models \phi_1 \wedge \phi_2 &\Leftrightarrow\quad & (x,t) \models \phi_1 \wedge (x,t) \models \phi_2 \\
(x,t) &\models \phi_1 U_{[a,b]} \phi_2 &\Leftrightarrow\quad & \exists t_1 \in [t+a, t+b]\ s.t. \\
& & & (x,t_1) \models \phi_2 \wedge \forall t_2 \in [t, t_1], \\
& & & (x,t_2) \models \phi_1 \\
(x,t) &\models \phi_1 \Rightarrow \phi_2 &\Leftrightarrow\quad & \text{if } (x,t) \models \phi_1 \text{ then } (x,t) \models \phi_2
\end{aligned} \quad (3)$$

To measure how well the STL formula is satisfied by the signal $x$, a quantitative semantics called robustness degree (RD) is used, mathematically denoted by $\rho$. If $\rho > 0$ the STL formula is satisfied, and $\rho < 0$ otherwise. If $\rho = 0$ the formula satisfaction is inconclusive [18] [19]. The robustness of $\phi$ with respect to signal $x$ at time $t$ is recursively defined as shown in (4).

$$\begin{aligned}
\rho^{\mu}(x,t) &= h(x) - c \\
\rho^{\sim \phi}(x,t) &= -\rho^{\phi}(x,t) \\
\rho^{\phi_1 \wedge \phi_2}(x,t) &= \min\left(\rho^{\phi_1}(x,t), \rho^{\phi_2}(x,t)\right) \\
\rho^{\phi_1 \vee \phi_2}(x,t) &= \max\left(\rho^{\phi_1}(x,t), \rho^{\phi_2}(x,t)\right) \\
\rho^{F_{[a,b]}\phi}(x,t) &= \max_{t' \in [t+a, t+b]} \left(\rho^{\phi}(x,t)\right) \\
\rho^{G_{[a,b]}\phi}(x,t) &= \min_{t' \in [t+a, t+b]} \left(\rho^{\phi}(x,t)\right) \\
\rho^{\phi_1 \Rightarrow \phi_2}(x,t) &= \max\left(-\rho^{\phi_1}(x,t), \rho^{\phi_2}(x,t)\right)
\end{aligned} \quad (4)$$

### B. Mathematical Model

Given the STL semantics, the following optimal control problem is formulated in (5) along a horizon $N_T$, maximizing RD subject to system dynamics. The objective is to find a control input sequence that satisfies a set of specifications.

$$\begin{aligned}
\max_{x,u} \rho(x) &- \alpha \sum_{k=1}^{N_T} \left[x_k^T Q x_k + u_k^T R u_k\right] \\
s.t. & \\
x_{k+1} &= f(x_k, u_k), \forall k \in \{0, 1, ..., N_T - 1\} \\
x_k &\in X, u_k \in U,
\end{aligned} \quad (5)$$

where $Q$ and $R$ are the weight matrices for state and control variables respectively, to form the Linear Quadratic Regulator (LQR) expression weighted with $\alpha$. Although this term penalizes the objective function, plays an important role to provide some stability on the state variables over the horizon.

Additionally, the computation of $\rho^{\phi}$ in (4) lacks of smoothness due to the max and min operators commonly used in RD

semantics. A smooth (infinitely differentiable) approximation is achieved by replacing the max and min operators in $\rho^\phi$ with the Log-Sum-Exponential (LSE) functions [20], as presented in (6) and (7) respectively.

$$\widetilde{\max}\left[(a_1, ..., a_m)\right]^T := \frac{1}{K} \log\left(\sum_{i=1}^{m} e^{ka_i}\right) \quad (6)$$

$$\widetilde{\min}\left[(a_1, ..., a_m)\right]^T := -\frac{1}{K} \log\left(\sum_{i=1}^{m} e^{-ka_i}\right) \quad (7)$$

Max and min operators approximation using LSE expressions is smooth, and its gradient can be analytically derived. As the parameter $K \to \infty$, the minimum and maximum values approach to their true values. Although the resulting mathematical model is still non-convex, employing convex approximations such as Taylor Series expansion and SCP allows to obtain a positive local maxima.

## III. PROBLEM FORMULATION AND IMPLEMENTATION OF MULTI-TASK LEARNING ON STL SPECIFICATIONS

A STL formula $\phi$ is composed by a predicate $\mu$ of the form $h(x) - c$, where $h(x)$ and $c$ describe the behavior of the state variable(s) in the task, in the standard form of $h(x) - c \geq 0$. The function $h(x)$ can be in terms of one or several state variables, depending on the application. Additionally, a time interval $[t_a, t_b]$ is also included, in which the specification $h(x) - c \geq 0$ should be met. The time interval $[t_a, t_b]$ can encompass the entire horizon $N_T$ or a segment of $N_T$. In general, the parameters $c$ and, $t_a$ and $t_b$ are the spatial and temporal parameters of the STL formula, respectively, formally expressed as $\phi_{[t_a, t_b]}$.

In this work, the philosophy behind the MTL is adopted to simultaneously train a model with multiple tasks, given that each task involves several specifications. Through spatial and temporal parameters perturbation within these tasks, the aim is to obtain a control and state variables sequence serving as a strong initializer to address other related tasks. This stage is refereed as "Learning stage", and the term "related" implies that the temporal and spatial parameters are affected by a similar level of perturbation in testing tasks compared with the learning tasks. Subsequently it is applied a phase known as "Testing stage", in which the solution of the tasks, i.e., the RD becomes positive, is achieved in only a few shots of SCP algorithm. This concept involves a quick online model adaptation to new tasks, as the learning is performed offline on a batch of related training tasks.

Algorithm 1 presents the procedure to generate $M$ tasks given $N$ specifications. Input data $ta$ and $tb$ are the lower and upper bounds respectively for the temporal parameters, and $c$ is the spatial parameter. These are included in the sets $ta = \{ta_1, \cdots, ta_N\}$, $tb = \{tb_1, \cdots, tb_N\}$ and $c = \{c_1, \cdots, c_N\}$. For each task, $N$ specifications are created by using the probability distribution of each temporal and spatial parameter, until $M$ tasks are generated.

Learning Stage is described in Algorithm 2. Using the function in Algorithm 1, $M_L$ tasks are generated and the average RD $\rho_{av}$ is computed following the semantics in (3)

**Algorithm 1** Task Generator

1: **function** TASKGENERATOR($M,N,ta,tb,c$)
2:     **for** $i = 1, \ldots, M$ **do**
3:         **for** $j = 1, \ldots, N$ **do**
4:             $ta'_{ij} \sim \mathcal{N}(ta_j, \sigma^2_{ta_j})$
5:             $tb'_{ij} \sim \mathcal{N}(tb_j, \sigma^2_{tb_j})$
6:             $c'_{ij} \sim \mathcal{N}(c_j, \sigma^2_{c_j})$
7:             $\phi_{i,j} \leftarrow ta_j, tb_j, c_j$
8:         **end for**
9:         Task$_i \leftarrow \{\phi_{i1}, \ldots, \phi_{iN}\}$
10:     **end for**
11:     Tasks $\leftarrow \{$Task$_1, \ldots,$ Task$_M\}$
12:     **return** Tasks
13: **end function**

and (4), and the LSE approximation in (6) and (7) for the max and min operators respectively. To solve the maximization problem for $\rho_{av}$, an affine approximation on the non-convex terms in $\rho_{av}$ is performed, previous to using SCP algorithm. Once the convergence is met in the SCP algorithm, i.e., the RD becomes positive and the trajectory of signal $x$ complies the specifications, the input $x_{\text{learn}}$ and $\mu_{\text{learn}}$ are obtained for testing stage.

**Algorithm 2** Learning Stage

1: **function** LEARNINGSTAGE($M_L, N, ta, tb, c$)
2:     Tasks$_{\text{learn}} \leftarrow$ TaskGenerator($M_L, N, ta, tb, c$)
3:     $\rho_{av} \leftarrow \frac{1}{M_L} \sum_{h=1}^{M_L} \rho(\text{Tasks}_{\text{learn}_h})$
4:     $\widehat{\rho_{av}} \leftarrow$ Affine($\rho_{av}$)
5:     $\rho^{\text{sol}}_{\text{learn}}, x_{\text{learn}}, \mu_{\text{learn}} \leftarrow$ SCP($\widehat{\rho_{av}}, A, B, x_0,$
6:                     $x_{\text{init}}, \mu_{\text{init}}, N_h$)
7:     **return** $x_{\text{learn}}, \mu_{\text{learn}}$
8: **end function**

In the testing stage described in Algorithm 3, the input data are the $x_{\text{learn}}$ and $\mu_{\text{learn}}$ previously obtained in Algorithm 2. Parameter $M_T$ is the number of testing tasks to generate in the testing stage. For each task, the affine approximation on the non-convex RD expression is performed. Once the convergence in SCP algorithm is reached, a solution for the testing task is obtained and stored in $\rho^{\text{sol}}_{\text{test}}$, in addition to the signal $x_{\text{test}}$, and the control sequence $\mu_{\text{test}}$.

**Algorithm 3** Testing Stage

1: **function** TESTINGSTAGE($x_{\text{learn}}, \mu_{\text{learn}}, M_T, N, ta, tb, c$)
2:     Tasks$_{\text{test}} \leftarrow$ TaskGenerator($M_T, N, ta, tb, c$)
3:     **for** $j = 1, \ldots, M_T$ **do**
4:         $\widehat{\rho}(\text{Tasks}_{\text{test}_j}) \leftarrow$ Affine($\rho(\text{Tasks}_{\text{test}_j})$)
5:         $x_h \leftarrow x_{\text{learn}}$
6:         $\mu_h \leftarrow \mu_{\text{learn}}$
7:         $\rho^{\text{sol}}_{\text{test}_j}, x_{\text{test}_j}, \mu_{\text{test}_j} \leftarrow$ SCP($\widehat{\rho}(\text{Tasks}_{\text{test}_j}),$
8:                     $A, B, x_0, x_h, \mu_h, N_h$)
9:     **end for**
10:     **return** $\rho^{\text{sol}}_{\text{test}}, x_{\text{test}}, \mu_{\text{test}}$
11: **end function**

As part of learning and testing stages, SCP algorithm is used to solve the non-convex mathematical model in (5). Algorithm 4 presents SCP pseudo-code. The input set is: convex approximation for the expression $\rho$, system dynamics $A$ and $B$, state variables initial point $x_0$, initial guess $x_{\text{init}}$ and $\mu_{\text{init}}$ for the signals and maximum number of iterations $N_h$ in the SCP. Since RD $\rho$ in its current form is non-convex due to negative LSE for the max operator, an affine approximation is applied using first-order Taylor Series, second-order Taylor series with positive semi-definite Hessian or inner convex approximations.

In this sense, the mathematical model in (5) is written in its linearized version as presented in (8).

$$\max_{x,u} \widehat{\rho}(x) - \sum_{k=1}^{N_T} \left[ x_k^T Q x_k + u_k^T R u_k \right]$$
$$s.t. \qquad (8)$$
$$x_{k+1} = \widehat{f}(x_k, u_k), \forall k \in \{0, 1, ..., N_T - 1\}$$
$$x_k \in X, u_k \in U,$$

where $\widehat{\rho}(x)$ is the linearized expression of the RD, which is obtained in (9) around a known operation point $x_{init}$.

$$\widehat{\rho}(x) = \rho(x_{init}) + \nabla \rho(x_{init})^T (x - x_{init})$$
$$+ \frac{1}{2} (x - x_{init})^T \nabla^2 \rho(x_{init}) (x - x_{init})$$
(9)

Expression $\widehat{f}(x_k, u_k)$ is the linearized expression for the system dynamics, which is addressed depending on the problem nature.

The initial guess $x_{\text{init}}$ and $\mu_{\text{init}}$ is given in the first iteration to linearize $\rho(x)$ and obtain $\widehat{\rho}(x)$. In the next iteration $h+1$, $x_h$ and $u_h$ obtained in the current iteration $h$ is used as nominal vector [21]. The update in the trust region is based on two performance metrics: relative error and relative decrease. The trust region shrinks/grows contingent upon the performance metrics comparison [22]. Convergence criteria is in terms of a maximum number of iterations $N_h$, moreover, SCP algorithm stops when RD of signal $x$ is positive and the signal meets the task specifications.

---

**Algorithm 4** Sequential Convex Programming

1: **function** SCP($\rho, A, B, x_0, x_{\text{init}}, \mu_{\text{init}}, N_h$)
2: $\quad x_h \leftarrow x_{\text{init}}$
3: $\quad \mu_h \leftarrow \mu_{\text{init}}$
4: $\quad$ **for** $h = 1, \ldots, N_h$ **do**
5: $\quad\quad \max_{x,\mu} \rho \leftarrow \text{solve}(8)$
6: $\quad\quad x_{h+1} \leftarrow x$
7: $\quad\quad \mu_{h+1} \leftarrow \mu$
8: $\quad\quad$ Update trust region
9: $\quad$ **end for**
10: $\quad$ **return** $\rho, x, \mu$
11: **end function**

---

The MTL overall framework is illustrated in Figure (1). The task generator (Algorithm 1), Sequential Convex Programming (Algorithm 4) and affine approximation modules are involved in both learning (Algorithm 2) and testing (Algorithm 3) stages.

## IV. CASE STUDIES AND RESULTS

For validation purposes, the MTL methodology is proved in two case studies. The first case corresponds to the Mass-Spring-Damper system where the mass position is controlled throughout the horizon time applying different STL specifications. In the second case, the Air Traffic Control (ATC) problem is addressed to generate a control sequence on a quadcopter, subject to preset rules in the STL framework. Learning and testing stages are run in Matlab R2023A using Gurobi solver in CVX modeling language.

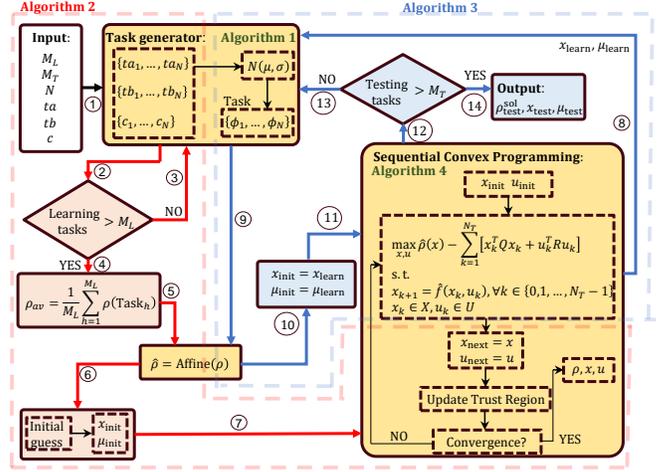

Fig. 1: Multi-task learning overall framework

### A. Mass-Spring-Damper system

The first scenario is the Mass-Spring-Damper system with the discretized dynamics shown in (10).

$$x_{k+1} = x_k + \left( \begin{bmatrix} 0 & 1 \\ \frac{-k_s}{m} & \frac{-b}{m} \end{bmatrix} x_k + \begin{bmatrix} 0 \\ \frac{1}{m} \end{bmatrix} \mu_k \right) \Delta_t \qquad (10)$$

where $x \in \mathbb{R}^2$ represents the variables $x_1$ and $x_2$: position and velocity of the mass, respectively. Control input is denoted by $\mu$. Mass, spring and damping constants are $m$, $k_s$ and $b$ respectively, and are set in $m = 1kg$, $k_s = 2N/m$ and $b = 0.2Ns/m$. Time step is chosen in $\Delta_t = 0.1s$. The base task with specifications is related with a desired behavior in the mass position $x_1$, along a horizon of 30 seconds and initial values for the state variables $[x_1 \quad x_2]_0 = [\pi \quad 2]$. The task is described as follows:

$$G_{[4,6]}(x_1 \geqslant 9) \wedge F_{[10,12]}(x_1 \leqslant -10) \wedge$$
$$G_{[16,18]}(x_1 \leqslant -12) \wedge F_{[22,24]}(x_1 \geqslant 13) \wedge \qquad (11)$$
$$G_{[28,30]}(x_1 \leqslant -15)$$

When the SCP is only applied on the task in (11), the mass position along the time horizon is shown in Figure 2a. The specifications are met considering the temporal parameters setting.

Figure (2b) presents the learning stage over 25 tasks randomly generated with a standard deviation of 1.1 in the temporal and spatial parameters. The average RD over all the tasks, becomes positive after 140 SCP iterations and stabilizes after 250 iterations. For the least and most complex tasks, were needed 119 and 207 SCP iterations respectively. A higher STD along the task generation, provides a more diverse set of tasks in the learning stage, furthermore, more SCP iterations are required to solve all the tasks. The solution obtained at this point is used as an initializer in the Testing stage.

For testing purposes, 10 tasks are randomly generated by applying perturbations using standard deviations in the temporal and spatial parameters. The behavior of the RD is plotted in Figure (2c) for two values of standard deviation, along with the corresponding dispersion across the tasks. Compared to the learning stage, it is observed that in the testing stage, the average RD becomes positive after a few SCP iterations: For perturbation levels of $STD = 2.5$ and $STD = 3.5$, were only required 9 and 20 SCP iterations respectively.

### B. ATC problem for autonomous quad-rotor

Air Traffic Control (ATC) problem involves complex rules to ensure safety in airspace use by aircraft. In the context of unmanned aerial vehicles, the ATC paradigm seeks to enable safe, efficient operation and route optimization in drone operations [23]. Flight constraints are generally related to altitude ranges for flying in certain areas and prohibited zones where aerial operations are not allowed, such as airports and military facilities [24]. In this work, the ATC problem is scaled to control a quad-rotor to reach a way-point and a final destination, i.e., addressing battery recharge and package delivery, following some altitude rules and flight avoidance of prohibited zones.

For simulation purposes, the linearized and discretized quad-rotor dynamics [25] is presented in (12a) and (12b):

$$x_{k+1} = Ax_k + B\mu_k \tag{12a}$$

$$A = \begin{bmatrix} 1 & 0 & 0 & 0 & 0 & 0 \\ 0 & 1 & 0 & 0 & 0 & 0 \\ 0 & 0 & 1 & 0 & 0 & 0 \\ 0.2 & 0 & 0 & 1 & 0 & 0 \\ 0 & 0.2 & 0 & 0 & 1 & 0 \\ 0 & 0 & 0.2 & 0 & 0 & 1 \end{bmatrix} \quad B = \begin{bmatrix} 1.96 & 0 & 0 \\ 0 & -1.96 & 0 \\ 0 & 0 & 0.4 \\ 0.196 & 0 & 0 \\ 0 & -0.196 & 0 \\ 0 & 0 & 0.04 \end{bmatrix} \tag{12b}$$

where $x \in \mathbb{R}^6$ represents the variables $x_1$, $x_2$ and $x_3$ corresponding to the velocities, and $x_4$, $x_5$ and $x_6$ the positions, in $x$, $y$ and $z$ coordinates respectively; and $\mu \in \mathbb{R}^3$ represents the control inputs $\mu_1$, $\mu_2$ and $\mu_3$ corresponding to the roll angle, pitch angle and thrust, respectively. Following the STL semantics, base task of autonomous quad-rotor in ATC framework is described in (13a) to (13e) along a 2 seconds horizon and initial values for the state variables, previously found using a LQR shot with some spatial location as reference.

$$G_{[0,2]}(q \in Zone_1 \Rightarrow 3 \leqslant x_6 \leqslant 7) \tag{13a}$$
$$G_{[0,2]}(q \in Zone_2 \Rightarrow 0 \leqslant x_6 \leqslant 4) \tag{13b}$$
$$G_{[0,2]}(x_4^2 + x_5^2 > 1.5^2) \tag{13c}$$
$$F_{[0,2]}((x_4 + 5)^2 + (x_5 + 2)^2 + (x_6 - 3)^2 \leqslant 0.5^2) \tag{13d}$$
$$F_{[0,2]}((x_4 + 2.5)^2 + (x_5 - 2.5)^2 + (x_6 - 1)^2 \leqslant 0.5^2) \tag{13e}$$

where $q$ is the quad-rotor position in $x$, $y$ and $z$ coordinates. The task involves (13a) $\wedge$ (13b) $\wedge$ (13c) $\wedge$ (13d) $\wedge$ (13e). Specifications (13a) and (13b) require that the altitude is within a lower and upper limit, depending on the zone.

During the time horizon, the quad-rotor is required to avoid a cylindrical unsafe zone, which is described in (13c). Way- and terminal points are represented as spheres in (13d) and (13e) respectively, that are eventually reached by the quad-rotor along $N_T$.

Considering the unsafe zone centered in the origin, and the task in (13a) to (13e), the trajectory performed by the quad-rotor is presented in Figure (2d), once the RD reaches a positive value in the SCP iterations. In some cases and depending on the level of perturbation assigned to the task, way- and terminal points might be visited in a different order, as there is no sequence of visit for these two points in the STL specifications.

Learning stage is performed for ATC problem considering 5 randomly generated tasks with a STD of 0.3 applied on spatial parameters expressed in (13a) and (13b). Radii of unsafe zone, way- and terminal spheres, are perturbed between a minimum and maximum value. Temporal parameters are not perturbed as there is no sequence based preference to perform specifications. Evolution of the maximum, average and minimum RD in this stage is depicted in Figure (2e), which become positive after 486, 542 and 1738 SCP iterations respectively.

Right at this point, the solution obtained in the learning stage is used as an initializer for the testing stage. Three values of standard deviation are considered to perturb the position of the way- and terminal points, which results in highly complex tasks. Radii are still modifying as proceeded in learning stage. Ten tasks are randomly generated, and for each perturbation level, the average RD is shown in Figure (2f) with dispersion across the tasks. Notice that the larger the perturbation level, the more SCP iterations are needed to obtain a positive average RD. In testing, SCP iterations required to meet the tasks are at most 4% compared with learning stage.

### V. CONCLUSIONS

A Multi-task learning (MTL) with Signal Temporal Logic (STL) framework is developed using Sequential Convex Programming (SCP). Approximations for min and max operators in the STL specifications are used to make affine the objective function in the mathematical model maximization, that overcomes the robustness degree of the overall STL task. This allows using SCP to solve the learning and testing stages of the MTL approach. For dynamical systems with temporal and spatial specifications, and considering different levels of perturbation in the STL specifications, the learning stage provided a strong initializer ($x_{learn}$ and $\mu_{learn}$) to solve related tasks in few SCP shots during the testing stage, even for highly perturbed tasks.

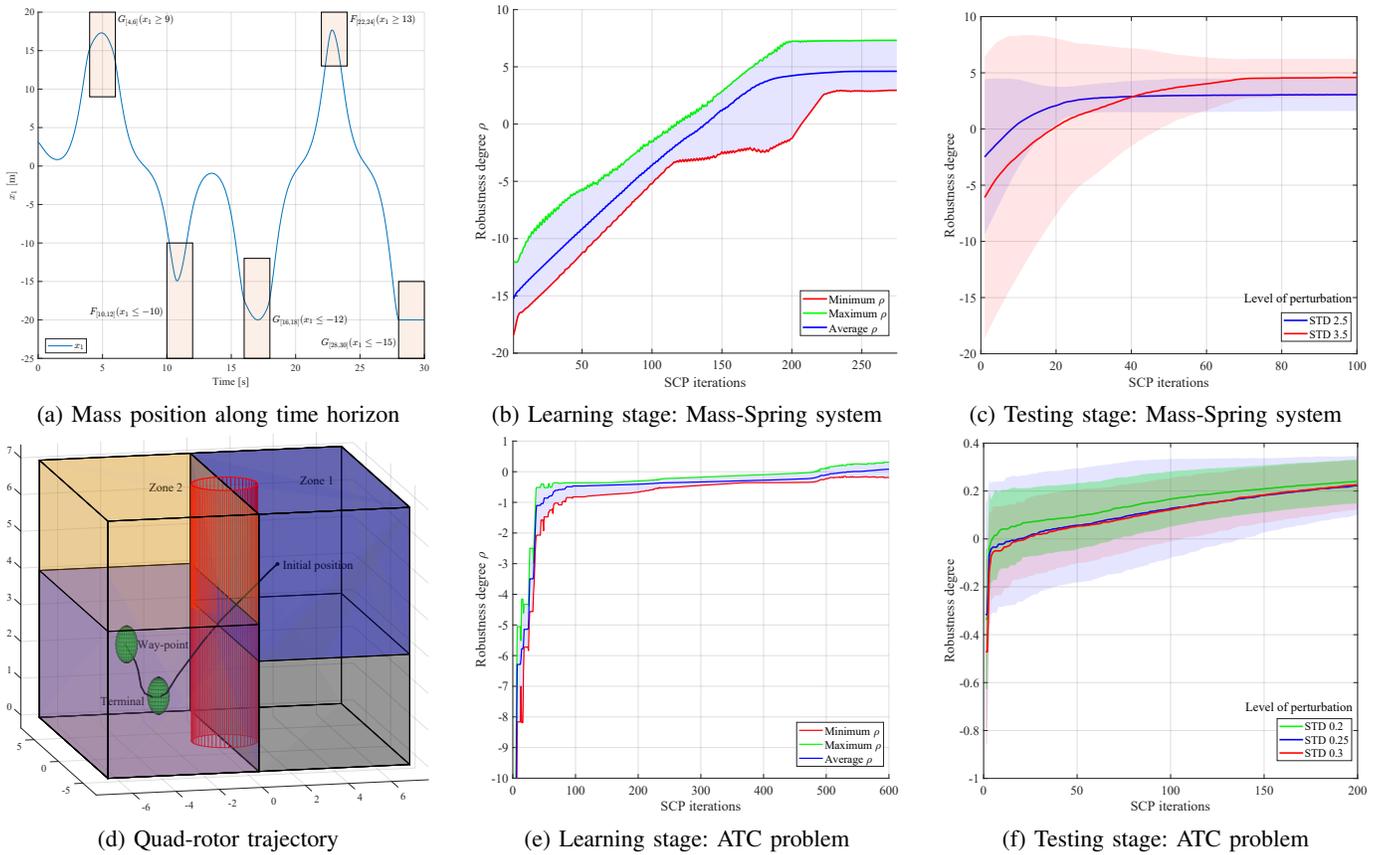

Fig. 2: Results for Mass-Spring-Damper system and ATC problem.